\documentclass[onecolumn,aps,prd,preprintnumbers,showpacs,superscriptaddress,nofootinbib,amsmath,amssymb,floats,floatfix,showkeys,notitlepage,longbibliography]{revtex4-1}
\usepackage{amsmath,amssymb}
\usepackage{graphicx,subfigure,epsfig}
\usepackage{hyperref}
\hypersetup{
  colorlinks=true,        
  linkcolor=blue,         
  citecolor=cyan,         
}
\usepackage{tablefootnote}
\usepackage{footmisc}
 \usepackage{booktabs}
 \usepackage{multirow}
 \usepackage[normalem]{ulem}
 \useunder{\uline}{\ul}{}
\usepackage{caption}
\usepackage{color}

\newcommand{\be}{\begin{equation}}
\newcommand{\ee}{\end{equation}}

 \pdfoutput=1 
 
 \usepackage[T1]{fontenc} 
 \usepackage{float}
 \usepackage{epstopdf}
 \usepackage{amsmath}
\makeatletter
\gdef\@fpheader{}
\makeatother

\begin{document}

\title{Topological dyonic black holes of massive gravity with generalized quasitopological electromagnetism}

\author{Askar Ali} 

\email{askarali@math.qau.edu.pk}
\affiliation{Department of Sciences and Humanities, National University of Computer and Emerging Sciences, Peshawar 25000, Pakistan}

\author{Ali \"Ovg\"un}
\email{ali.ovgun@emu.edu.tr}
\affiliation{Physics Department, Eastern Mediterranean University, Famagusta, 99628 North Cyprus via Mersin 10, Turkiye.}

\date{\today}

\begin{abstract}In this paper we investigate new dyonic black holes of massive gravity sourced by generalized quasitopological electromagnetism in arbitrary dimensions. We begin by deriving the exact solution to the field equations defining these black holes and look at how graviton's mass, dimensionality parameter, and quasitopological electromagnetic field affect the horizon structure of anti-de Sitter dyonic black holes. We also explore the asymptotic behaviour of the curvature invariants at both the origin and infinity to analyze the geometric structure of the resultant black holes. We also compute the conserved and thermodynamic quantities of these dyonic black holes with the help of established techniques and known formulas. After investigating the relevancy of first law, we look at how various parameters influence the local thermodynamic stability of resultant black hole solution. We also examine how thermal fluctuations affect the local stability of dyonic black holes in massive gravity. Finally, we study the shadow cast of the black hole.
\end{abstract}


\maketitle

\section{Introduction}
\label{sec:intro}
Einstein's theory of gravity (ETG) is a relativistic model that describes the gravitational field. In this configuration, it is presumed that the graviton has no mass. The ability to construct a self-consistent notion of gravity in the case where the graviton has mass is an expected question. Recent Ligo experiment observations also imply that the graviton has a nonzero mass \cite{1a}. Additionally, it is claimed that the Hubble scale's massive graviton may be the root cause for the Universe's accelerated expansion \cite{1b,1c}. A family of nonlinear massive gravity theories were developed in Refs. \cite{2b,2c,2d}. Note that these theories do not include ghost fields \cite{2e}. In the four dimensional massive gravity, Vegh \cite{2g} discovered a nontrivial black hole solution with a negative cosmological constant. This solution has also a Ricci flat horizon \cite{2h}. The graviton's mass was later discovered to have the same act as lattice in case of holographic conductor model: typically, the conductivity displays a Drude peak that resembles a delta function when the graviton's mass vanishes. In Refs. \cite{3b,3c,3d,3e,3f}, some holographic repercussions of the impact of the nonzero graviton mass were also explored. The metal-insulator transition is another intriguing scenario that gives credibility to the massive gravities \cite{4a}. The higher dimensional Vegh's black holes of massive gravity were explored in Ref. \cite{4b}.  

Massive gravities should be recognized as requiring not just the dynamical metric $g_{\mu\nu}$ but also the fiducial reference metric $f_{\mu\nu}$ and non-derivative potentials $\Xi_i$'s. It is quite intriguing that a new class of massive gravities can be developed when an appropriate choice for the metric $f_{\mu\nu}$ is made \cite{17a}. For numerous fiducial metrics, including the Minkowskian and singular references metrics \cite{18a,19a,20a,21a,22a}, the dRGT version of massive gravity \cite{2c} emerges out to be a ghost-free. Studies of various cosmological scenarios in massive gravity revealed that they are reliant on the selection of the Minkowskian reference metrics \cite{21a,22a,23a,24a,25a}. In Refs. \cite{26a,27a,28a,29a,30a,30b,30c,30d,30e,30e1,30e2,30f,30g,30h}, different black holes of the massive gravity and their physical properties have recently been examined. Among these, the anti-de Sitter black holes play a significant role in various usages of gauge/gravity duality \cite{2h} and black hole chemistry \cite{31a}. 

Recently Liu et al. introduced the quasitopological electromagnetism within the framework of Abelian gauge theories \cite{32a}. They altered the conventional Maxwellian model by incorporating new terms into the Lagrangian. Remember that these terms depend on the metric tensor and electromagnetic $2$ -form $F_{[2]}=dA_{[1]}$. These involve polynomials such as
\begin{equation}
V_{[2s]}=F_{[2]}\wedge F_{[2]}\wedge\cdots\wedge F_{[2]}. \label{5a}
\end{equation}        
The above polynomials seem like the Pontryagin densities, and their integrals in even dimensions are truly topological. The dynamics of the system can, however, be affected when these $2s$-forms are adopted for arbitrary dimensions. One can attained this by using the squared norm
\begin{equation}
U^{(s)}_{[d]}\sim\left|V_{[2s]}\right|^2\sim V_{[2s]}\wedge \ast V_{[2s]}. \label{6a} 
\end{equation} 
Remember that the situation with $s=1$ in Eq. (\ref{6a}) resembles the standard kinetic term of Maxwell's model. The contributions of these invariants to the field equations are often non-vanishing. The following two factors have led to the designation of this model as quasitopological electromagnetism. First, as a result of the forms $V_{[2s]}$, i.e. the component pieces of this model, have topological origin. Second, the spectra of purely electric and magnetic solutions are in agreement with the spectra of the solutions that correspond in Maxwell's theory. Investigating dyons, though, reveals amazing things. Recently, the further development in this regard has been examined as well so that the Abelian gauge field $A_{[1]}$ and the higher-rank $(s-1)$-form field $B_{[s-1]}$ are both taken into account \cite{32b}. By using the relation $H_{[s]}=dB_{[s-1]}$, one can defined the field strength that relates to the value of $B_{[s-1]}$. This generalized model may have several physical premises according to the specifics of the situation. The fundamental Kalb-Ramond $2$-form $B_{[2]}$, the Ramond-Ramond $s$-forms, or the $3$-forms of the eleventh dimensional super-gravity are some examples of higher rank fields formulating in string theory. Additionally, it has been established that more prevalent black holes can develop when the Maxwell field is coupled to $s$-forms \cite{32b}. Recently, black hole configurations with scalar hair in the background of Lovelock gravity and generalized quasitopological electromagnetism have also been addressed \cite{32c}. In this study, we are eager to investigate dyonic black holes of massive gravity with generalized quasitopological electromagnetism.

A striking illustration of this concept can be found in the groundbreaking images of the space-time surrounding supermassive black holes, captured by the Event Horizon Telescope team \cite{EventHorizonTelescope:2019dse}. This remarkable journey into unraveling the enigmas of deep spacetime commenced in 2019 with the historic unveiling of the first image of the compact object M87*. Continuing on this trajectory, the EHT team has recently astounded us once again by documenting the inaugural visuals of another supermassive compact object situated at the heart of our Milky Way galaxy—Sagittarius A* (Sgr A*) \cite{EventHorizonTelescope:2022apq}. Subsequently, numerous works have emerged in this domain, reflecting the ongoing exploration and research in this captivating field \cite{Ovgun:2018tua,Ovgun:2020gjz,Ovgun:2019jdo,Yang:2023tip,Okyay:2021nnh,Yang:2023agi,Kumaran:2023brp,Lambiase:2023hng,Uniyal:2022vdu,Uniyal:2023inx,Ovgun:2023ego,Atamurotov:2022knb,Kumaran:2022soh,Pantig:2022gih,Pantig:2022qak,Mustafa:2022xod,Rayimbaev:2022hca,Gullu:2020qzu,Pantig:2022ely,Kuang:2022xjp,Cimdiker:2021cpz,Cimdiker:2023zdi,Pulice:2023dqw,Pantig:2022sjb,Pantig:2022toh,Pantig:2021zqe,Pantig:2022whj,Feng:2022evy,Banerjee:2022jog,Falcke:1999pj,Xu:2018mkl,Hou:2018bar,Chakhchi:2022fls,Tsukamoto:2014tja,Tsukamoto:2017fxq,Shaikh:2019fpu,Shaikh:2018kfv,Kasuya:2021cpk,Fathi:2022ntj,Belhaj:2020kwv,Belhaj:2020okh,Vagnozzi:2022moj,Chen:2022nbb,Roy:2021uye,Vagnozzi:2020quf,Bambi:2019tjh,Bambi:2010hf,Zakharov:2014lqa,Zakharov:2005ek,Zakharov:2011zz,Zakharov:2018syb,Zakharov:2021gbg,KumarWalia:2022aop,Konoplya:2021slg,Churilova:2021tgn,Konoplya:2019fpy,Konoplya:2019goy,Younsi:2016azx,Khodadi:2021gbc,Xavier:2023exm,Parbin:2023zik,Mustafa:2023vvt,Capozziello:2023rfv,Qin:2023nog,Mustafa:2022fxn,Pugliese:2022oes,Konoplya:2022hbl,Konoplya:2019sns,Bhandari:2021dsh,Contreras:2020kgy,Contreras:2019cmf,Contreras:2019nih,Herdeiro:2021lwl,Sengo:2022jif,Junior:2021svb,Lima:2021las,Cunha:2018cof,Cunha:2018gql,Cunha:2015yba,Guerrero:2021ues,Guerrero:2021pxt,Sharif:2016znp,Khodadi:2022ulo,Meng:2023wgi,Wang:2023vcv,Li:2023djs,Kuang:2022ojj,Meng:2022kjs,Anjum:2023axh,Islam:2022wck,Ghosh:2022kit,Ghosh:2020spb,Abdujabbarov:2016hnw,Kumar:2020owy,Papnoi:2014aaa,Kumar:2018ple,Toshmatov:2014nya,Atamurotov:2013sca,Abdujabbarov:2015xqa,Atamurotov:2015nra}.

The outline of the paper is as follows. In Section 2, we coupled massive gravity to generalized quasitopological electromagnetism and used the variational principle to construct the updated equations of motion. In this configuration, a new topological dyonic black hole solution of massive gravity is generated. Section 3 addresses the thermodynamic properties of the resulting solutions. In Section 4, we look into how thermal fluctuations alter the physical quantities of the black hole.  In Section 5,  we study the shadow cast of the black hole. Finally, Section 6 provides some concluding thoughts to wrap up the paper.

\section{Topological dyonic black holes} 
The action of the massive gravity with cosmological constant $\Lambda$ and generalized quasitopological electromagnetism can be expressed as  
\begin{equation}
\mathcal{I}=\int d^dx\sqrt{-g}\bigg[R+m^2\sum_{i=1}^{4}a_i\Xi_i(\textbf{g},\textbf{f})-2\Lambda+\mathfrak{L}_{QT}\bigg],
\label{1}
\end{equation}
in which $\mathfrak{L}_{QT}$ stands for the Lagrangian density of extended quasitopological electromagnetic field and $R$ signifies the Ricci scalar. In addition, $m$ denotes the mass of graviton, $a_i$'s are the coupling constants, and $\Xi$'s are signifying the non-derivative potentials. Notably, these potentials are symmetric polynomials that specify the eigenvalues of the $d\times d$ matrix $\zeta^{\mu}_{\nu}=\sqrt{g^{\mu\alpha}f_{\alpha\nu}}$. These potentials can be addressed as
\begin{eqnarray}\begin{split}
\Xi_1=[\zeta], \Xi_2=[\zeta]^2-[\zeta^2], 
\Xi_3=[\zeta]^3-3[\zeta][\zeta^2]+2[\zeta^3],\\
\Xi_4=[\zeta]^4-6[\zeta^2][\zeta]^2+8[\zeta^3][\zeta]+3[\zeta^2]^2-6[\zeta^4]. \label{2}\end{split}
\end{eqnarray}

\begin{figure}[h]
    	\centering
    	\includegraphics[width=0.8\textwidth]{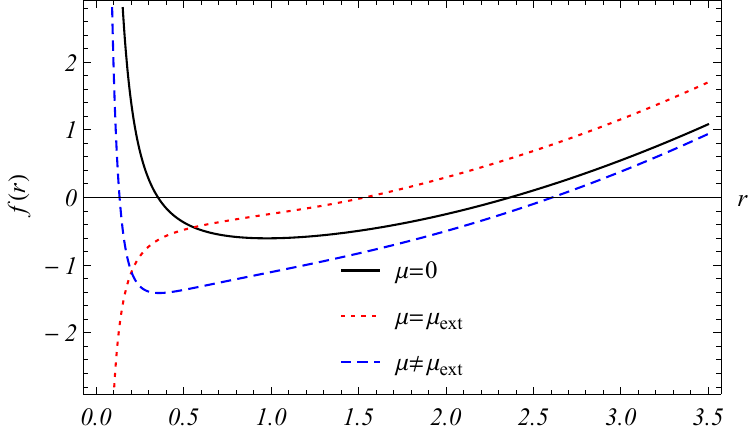}
    	\caption{Plot of the resultant solution $f(r)$ (Eq. (\ref{17})) for various values of geometric mass. The other values have been taken as $d=4$, $\Theta=-1$, $\Lambda=-0.5$, $\nu=0.5$, $m=0.05$, $e=1$, $a_1=1$, $a_2=2$, $a_3=1$, $a_4=2$, $\beta=1$, and $\chi=1$.}\label{skr1A}
    \end{figure}

\begin{figure}[h]
    	\centering
    	\includegraphics[width=0.8\textwidth]{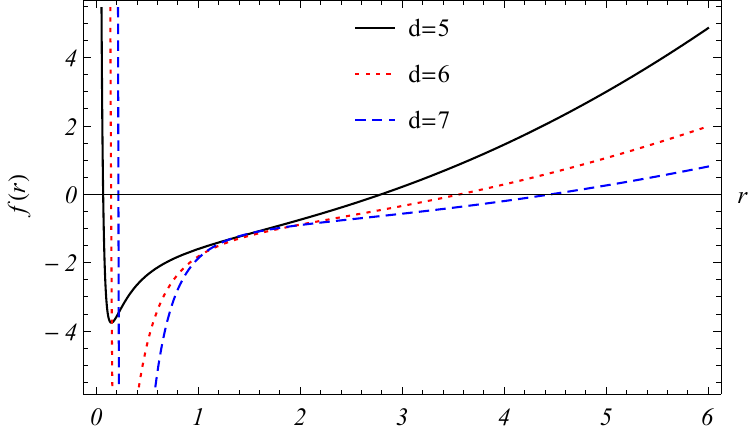}
    	\caption{Plot of the resultant solution $f(r)$ (Eq. (\ref{17})) in various dimensions. The other values have been taken as $\mu=1$, $\Theta=-1$, $\Lambda=-0.5$, $\nu=0.5$, $m=0.05$, $e=1$, $a_1=1$, $a_2=2$, $a_3=1$, $a_4=2$, $\beta=1$, and $\chi=1$.}\label{skr1B}
    \end{figure}

\begin{figure}[h]
   	\centering
   	\includegraphics[width=0.8\textwidth]{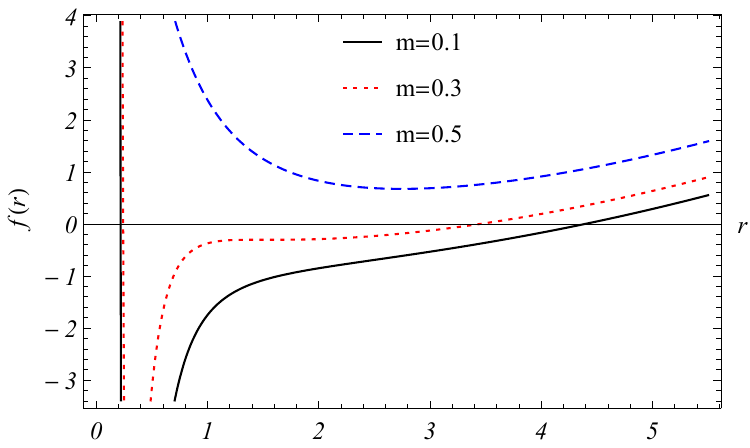}
   	\caption{Dependence of the resultant solution $f(r)$ (Eq. (\ref{17})) on the graviton's mass. The other values have been taken as $d=6$, $\Theta=-1$, $\Lambda=-0.5$, $\nu=0.5$, $\mu=1$, $e=1$, $a_1=1$, $a_2=2$, $a_3=1$, $a_4=2$, $\beta=1$, and $\chi=1$.}\label{skr1C}
   \end{figure}

\begin{figure}[h]
	\centering
	\includegraphics[width=0.8\textwidth]{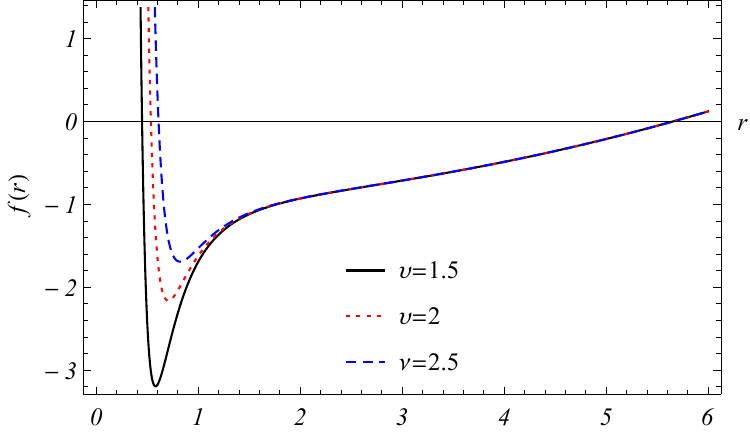}
	\caption{Impact of the magnetic charge on the resultant solution $f(r)$ (Eq. (\ref{17})). The other values have been taken as $d=6$, $\Theta=-1$, $\Lambda=-0.3$, $\mu=1$, $m=0.05$, $e=1$, $a_1=1$, $a_2=2$, $a_3=1$, $a_4=2$, $\beta=1$, and $\chi=1$.}\label{skr1D}
\end{figure}

\begin{figure}[h]
	\centering
	\includegraphics[width=0.8\textwidth]{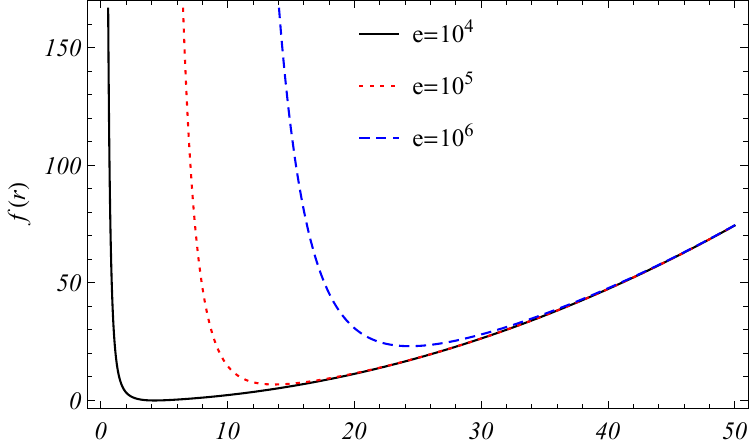}
	\caption{Behaviour of the resultant solution $f(r)$ (Eq. (\ref{17})) for various values of the electric charge. The other values have been taken as $d=6$, $\Theta=-1$, $\Lambda=-0.3$, $\nu=2.5$, $m=0.05$, $\mu=1$, $a_1=1$, $a_2=2$, $a_3=1$, $a_4=2$, $\beta=1$, and $\chi=1$.}\label{skr1E}
\end{figure}

The Lagrangian density of the generalized quasitopological electromagnetism \cite{32b} is stated as
\begin{equation}
\mathfrak{L}_{QT}=-\frac{1}{4} F_{\rho\sigma}F^{\rho\sigma}-\frac{1}{2s!}H_{\nu_1\nu_2\cdots\nu_s}H^{\nu_1\nu_2\cdots\nu_s}-\chi\mathfrak{L}_{int},\label{3}
\end{equation}
in which $\chi$ is the coupling constant and $\mathfrak{L}_{int}$ designates the interaction term, which is expressed as
\begin{equation}
\mathfrak{L}_{int}=\delta^{\rho_1\cdots\rho_d}_{\sigma_1\cdots\sigma_d}F_{\rho_1\rho_2}H_{\rho_3\cdots\rho_d}F^{\sigma_1\sigma_2}H^{\sigma_3\cdots\sigma_d}.\label{4}
\end{equation}
Remember that the following forms are taken into consideration for the formation of above Eq. (\ref{4}), i.e. 
\begin{align}\begin{split}
&F_{\rho\sigma}\sim h'(r)\delta^{x^0x^1}_{\rho\sigma},\\&
H_{\nu_1\nu_2\cdots\nu_s}\sim \delta^{x^2\cdots x^d}_{\nu_1\nu_2\cdots\nu_s},\label{5}\end{split}
\end{align} 
with $s=d-2$, while $F_{\mu\nu}$ and $ H_{\nu_1\nu_2\cdots\nu_s}$ are purely electric and purely magnetic, respectively. It is possible to obtain the field equations characterizing quasitopological electromagnetism through the use of Eqs. (\ref{3})-(\ref{5}) in Eq. (\ref{1}) as follows:
\begin{equation}
\nabla_{\gamma}F^{\gamma\rho}-4\chi\delta^{\rho\gamma\nu_1\cdots\nu_s}_{\sigma_1\sigma_2\cdots\sigma_d}H_{\nu_1\cdots\nu_s}\nabla_{\gamma}\big(F^{\sigma_1\sigma_2}H^{\sigma_3\cdots\sigma_d}\big)=0,\label{6}
\end{equation} 
and
\begin{equation}
\nabla_{\rho}H^{\rho\nu_1\cdots\nu_{s-1}}+2\chi s!\delta^{\rho\gamma\beta\nu_1\cdots\nu_{s-1}}_{\sigma_1\cdots\sigma_d}F_{\rho\gamma}\nabla_{\beta}\big(F^{\sigma_1\sigma_2}H^{\sigma_3\cdots\sigma_d}\big)=0.\label{7}
\end{equation}
By varying Eq. (\ref{1}) with regards to $g_{\mu\nu}$, the equations of motion of massive gravity sourced by quasitopological electromagnetic field can be derived as
\begin{equation}
R_{\mu\nu}-\frac{1}{2}g_{\mu\nu}R+\Lambda g_{\mu\nu}+m^2X_{\mu\nu}=\mathfrak{T}^{(QT)}_{\mu\nu},
\label{8}
\end{equation}
with 
\begin{eqnarray}
\label{8a}
\begin{split}
X_{\mu\nu}&=-\frac{a_1}{2}\bigg(\Xi_1g_{\mu\nu}-\zeta_{\mu\nu}\bigg)-\frac{a_2}{2}\bigg(\Xi_2g_{\mu\nu}-2\Xi_1\zeta_{\mu\nu}+2\zeta^2_{\mu\nu}\bigg)-\frac{a_3}{2}\\&\times\bigg(\Xi_3g_{\mu\nu}-3\Xi_2\zeta_{\mu\nu}+6\Xi_1\zeta^2_{\mu\nu}-6\zeta^3_{\mu\nu}\bigg)-\frac{a_4}{2}\bigg(\Xi_4g_{\mu\nu}-4\Xi_3\zeta_{\mu\nu}\\&+2\Xi_2\zeta^2_{\mu\nu}-24\Xi_1\zeta^3_{\mu\nu}+24\zeta^4_{\mu\nu}\bigg).
\end{split}
\end{eqnarray}
 Furthermore, the energy-momentum tensor of the quasitopological electromagnetic field is labeled by $\mathfrak{T}^{(QT)}_{\mu\nu}$ and is specified as
\begin{equation}\begin{split}
\mathfrak{T}^{(QT)}_{\alpha\beta}&=F_{\alpha\rho}F^{\rho}_{\beta}-\frac{1}{4}g_{\alpha\beta}F_{\rho\sigma}F^{\rho\sigma}+\frac{1}{2(s-1)!}H_{\alpha\nu_1\cdots\nu_{s-1}}H^{\nu_1\cdots\nu_{s-1}}_{\beta}\\&-\frac{1}{2(s!)^2}\delta^{\nu_1\cdots\nu_s\rho}_{\sigma_1\cdots\sigma_s(\alpha}g_{\beta)\rho}H_{\nu_1\cdots\nu_s}H^{\sigma_1\cdots\sigma_s}+\chi g_{\alpha\beta}\mathfrak{L}_{int}.
\label{9}\end{split}
\end{equation}
Here, we are attempting to figure out the dyonic black hole solution of the field equations (\ref{8}) by using the generalized model of quasitopological electromagnetism. To execute this, we are incorporating the following metric ansatz
\begin{equation}
ds^2=-f(r)dt^2+\frac{dr^2}{f(r)}+r^2d\Upsilon^{2}_{d_2}.
\label{10}
\end{equation}
Remember that we are employing $d_j=d-j$ for simplicity. Meanwhile, $d\Upsilon^{2}_{d_2}$ specifies the metric of the $d_2$-dimensional submanifold with constant curvature $d_2d_3\Theta$, and can be presented as
\begin{equation}\begin{split}
d\Upsilon^2_{d_2}=\left\{ \begin{array}{rcl}
d\theta_1^2+\sum_{j=2}^{d-2}\prod_{l=1}^{j-1}\sin^2\theta_l d\theta_j^2, & 
& \Theta=1, \\d\theta_1^2+\sinh^2\theta_1d\theta_2^2+\sinh^2\theta_1\sum_{j=3}^{d-2}\prod_{l=2}^{j-1}\sin^2\theta_ld\theta^2_j, &  & \Theta=-1,\\\sum_{j=1}^{d-2}d\phi_j^2, & & \Theta=0,
\end{array}\right.\label{11}
\end{split}
\end{equation}
Remember that this sub-manifold is embellished with a magnetic field that is proportionated to its intrinsic volume form such that
\begin{equation}
H_{\nu_1\nu_2\cdots\nu_m}=\nu\sqrt{\Sigma}\delta^{x^1\cdots x^m}_{\nu_1\cdots\nu_m}.\label{12}
\end{equation} 
Similarly, for the purely electric case, we can write
\begin{equation}
F_{\alpha\beta}=h'(r)\delta^{tr}_{\alpha\beta}, \label{13}
\end{equation}
where prime refers to the differentiation with respect to $r$. Thereby, through the usage of Eqs. (\ref{12}) and (\ref{13}) one finds
\begin{equation}
r^{2d_2}\big(d_2h'(r)+rh''(r)\big)-8\chi(d_2!)^2q^2\big(d_2h'(r)-rh''(r)\big)=0.\label{14}
\end{equation} 
One may integrate the above equation to get
\begin{equation}
h'(r)=\frac{e r^{d_2}}{r^{2d_2}+8\chi(d_2!)^2\nu^2}.\label{15}
\end{equation}
Here the integration constants $\nu$ and $e$ are respectively related to magnetic and electric charges. Moreover,  Eq. (\ref{15}) specifies the screening of the electric field driven by the interaction with the magnetic part. Since the spacetime is presupposed to have a $d$-dimensional metric (\ref{10}), it is beneficial to set up the extra dimensional reference metric as $f_{\mu\nu}=Diag(0,0,\beta^2h_{ij})$, with $\beta>0$ and $h_{ij}$ conveying the sector that refers to $d_2$-dimensional submanifold in Eq. (\ref{11}). It is simple to estimate $\Xi_i$'s in the form $\Xi_k=\frac{\beta^k}{r^k}\prod_{j=2}^{k+1}d_j$ by employing this extended reference metric \cite{4b,33a}. As a result, by substituting Eq. (\ref{10}), fiducial metric $f_{\mu\nu}$, and energy-momentum tensor (\ref{9}) with Eqs. (\ref{12}) and (\ref{15}) in the gravitational field equations (\ref{6}), one may extract the following solution
      \begin{equation}\begin{split}
     f(r)&=\Theta-\frac{\mu}{r^{d_3}}-\frac{2\Lambda r^2}{d_1d_2}+m^2\bigg(\frac{a_1\beta r}{d_2}+a_2\beta^2+\frac{a_3\beta^3d_3}{r}+\frac{a_4\beta^4d_3d_4}{r^2}\bigg)\\&+\frac{e^2}{2d_2d_3r^{2d_3}} F_1\bigg(\bigg[1,\frac{d_3}{2d_2}\bigg],\bigg[\frac{3d-7}{2d_2}\bigg],\frac{-8\chi \nu^2(\Gamma(d_1))^2}{r^{2d_2}}\bigg)+\frac{\nu^2}{2d_2d_3r^{2d_3}},\label{17}\end{split}
     \end{equation}
      where the integration constant $\mu$ refers to the geometric mass of black hole and $F_1$ denotes the hypergeometric function. In four dimensional spacetimes, the resultant solution (\ref{17}) yields a simpler form
      \begin{equation}\begin{split}
          f(r)&=\Theta-\frac{\mu}{r}-\frac{\Lambda r^2}{3}+m^2\bigg(\frac{a_1\beta r}{2}+a_2\beta^2+\frac{a_3\beta^3}{r}\bigg)+\frac{\nu^2}{4r^2}\\&+\frac{e^2}{4r^2}F_1\bigg(\bigg[1,\frac{1}{4}\bigg],\bigg[\frac{5}{4}\bigg],\frac{-32\chi \nu^2}{r^{4}}\bigg),\label{17a} \end{split}
      \end{equation}
      which converts into four dimensional Reissner-Nordstrom type solution of massive gravity for large values of $r$, i.e., 
      \begin{equation}\begin{split}
          f(r)&=\Theta-\frac{\mu}{r}-\frac{\Lambda r^2}{3}+m^2\bigg(\frac{a_1\beta r}{2}+a_2\beta^2+\frac{a_3\beta^3}{r}\bigg)+\frac{\nu^2+e^2}{4r^2}.\label{17b} \end{split}
      \end{equation}
      The dependence of the resultant solution (\ref{17}) on the geometric mass $\mu$ is shown in Fig. \ref{skr1A}. The equation $f(r)=0$ can be utilized to figure out the inner and outer horizons. The positions of horizons correspond to those values of $r$ at which the curve connected to dyonic solution (\ref{17}) crosses the horizontal axis. \textcolor{black}{Moreover, the event horizon $r_{ext}$ and mass $\mu_{ext}$ of an extreme black hole can be worked out if one solves the equations $f(r)=0$ and $f'(r)=0$, simultaneously \cite{33b}. Hence, it is simple to obtain the geometric mass of an extreme black hole as follows:
      \begin{equation}\begin{split}
      \mu_{ext}&=\frac{r_{ext}^{d_3}}{d_1}\bigg[2\Theta+m^2\bigg(\frac{a_1\beta r_{ext}}{d_2}+2a_2\beta^2+\frac{3a_3\beta^3}{r_{ext}}+\frac{4a_4\beta^4}{r_{ext}^2}\bigg)+\frac{\nu^2}{d_3r_{ext}^{2d_3}}\\&+\frac{e^2}{d_3r_{ext}^{2d_3}} F_1\bigg(\bigg[1,\frac{d_3}{2d_2}\bigg],\bigg[\frac{3d-7}{2d_2}\bigg],\frac{-8\chi \nu^2(\Gamma(d_1))^2}{r_{ext}^{2d_2}}\bigg)\\&+\frac{e^2r_{ext}^2}{2d_2\big(r_{ext}^{2d_2}+8\chi\nu^2(\Gamma(d_1))^2\big)}\bigg].\label{17c}\end{split}
      \end{equation}
       Note that the event horizon of extreme black hole satisfies the following equation
       \begin{equation}\begin{split}
       &d_3\Theta-\frac{2\Lambda r_{ext}^2}{d_2}+m^2\bigg(a_1\beta r_{ext}+d_3a_2\beta^2+\frac{d_3d_4a_3\beta^3}{r_{ext}}+\frac{d_3d_4d_5a_4\beta^4}{r_{ext}^2}\bigg)\\&-\frac{\nu^2}{2d_2r_{ext}^{2d_3}}-\frac{e^2r_{ext}^2}{2d_2\big(r_{ext}^{2d_2}+8\chi\nu^2(\Gamma(d_1))^2\big)}=0.\label{17d}\end{split}
       \end{equation}
       Fig. \ref{skr1A} shows that when geometric mass attains its extreme value $\mu_{ext}$, the resultant dyonic black solution has only one event horizon. The interesting scenario arises when $\mu\neq\mu_{ext}$. In this case, the dyonic black holes of massive gravity have both inner Cauchy and outer event horizons even when $\mu=0$. This interesting behaviour of the metric function is due to the assumption of massive graviton in the action function.} Fig. \ref{skr1B} exhibits the behaviour of the resultant solution in various dimensions. Notably, the horizon structure of the resultant dyonic solution is also influenced by the dimensionality parameter. It is demonstrated that the higher dimensional black holes have larger Cauchy and event horizons than the lower dimensional objects. \textcolor{black}{In addition, the impact of massive graviton on the resultant solution is demonstrated in Fig. \ref{skr1C}. We have noted earlier that the solution (\ref{17}) describes dyonic black holes with inner and outer horizons for any non-negative value of the black hole's  geometric mass such that $\mu\neq\mu_{ext}$. However, there also exists a critical value $m_c$ of graviton's mass such that the resultant solution describes a black hole with two horizons when the mass of graviton is lesser than this critical value, an extreme black hole having one horizon when the mass of graviton takes the value $m_c$, and naked singularity when graviton's mass exceeds the value $m_c$.} Similarly, electric and magnetic charges have also unavoidable effects on the behaviour of our resultant solution (\ref{17}). Figs. \ref{skr1D} and \ref{skr1E} demonstrate that the radius of the Cauchy horizon grows when the magnitude of these charges increases, however, the location of the outer horizon remains unchanged.  
     The Ricci and Kretschmann invariants that correlate with Eq. (\ref{10}) are stated as
     \begin{eqnarray}\begin{split}
     R(r)&=(d^2-5d+6)\frac{(\Theta-f(r))}{r^2}-\frac{d^2f}{dr^2}-\frac{2d_2}{r}\frac{df}{dr},\label{18}\end{split}
     \end{eqnarray}
     and
     \begin{eqnarray}\begin{split}
     K(r)&=2d_2d_3\frac{(k-f(r))^2}{r^4}-\bigg(\frac{d^2f}{dr^2}\bigg)^2+\frac{2d_2}{r^2}\bigg(\frac{df}{dr}\bigg)^2.\label{19}\end{split}
     \end{eqnarray}
     Hence, using the resultant solution (\ref{17}) in Eqs. (\ref{18})-(\ref{19}), one can show that
     \begin{align}\begin{split}
     &\lim_{r\rightarrow\infty} R(r)=\frac{2d}{d_2}\Lambda,\\&
     \lim_{r\rightarrow\infty} K(r)=\frac{8d}{d_1d_2^2}\Lambda^2.\label{20}\end{split}
     \end{align}
Similarly, it can also be shown that
\begin{align}\begin{split}
&\lim_{r\rightarrow 0} R(r)\rightarrow\infty,\\&
\lim_{r\rightarrow 0} K(r)\rightarrow\infty.\label{21}\end{split}
\end{align}
The irregular behaviour of these curvature invariants at $r=0$ indicates the presence of a true curvature singularity at this position. Hence, Eqs. (\ref{20}) shows that the metric function (\ref{17}) describes the new family of non-asymptotically flat black holes of massive gravity sourced by generalized quasitopological electromagnetism. It should also be remembered that the metric function (\ref{17}) reduces to the higher dimensional dyonic solution of ETG \cite{32b} when one puts $m=0$.
\section{Thermodynamics of dyonic black holes}

Now we seek to figure out the thermodynamic and conserved quantities of the resultant dyonic black holes (\ref{17}). From $f(r_+)=0$, it is feasible to exhibit the finite mass as
\begin{eqnarray}\begin{split}
M&=\frac{d_2\mu\Sigma_{d_2}}{2}=\frac{d_2\Sigma_{d_2}}{2}\bigg[\Theta r_+^{d_3}-\frac{2\Lambda}{d_1d_2}r_+^{d_1}+\frac{\nu^2}{2d_2d_3r_+^{d_3}}\\&+\frac{e^2}{2d_2d_3r_+^{d_3}}F_1\bigg(\bigg[1,\frac{d_3}{2d_2}\bigg],\bigg[\frac{3d_2-1}{2d_2}\bigg],\frac{-8\chi \nu^2(\Gamma(d_1))^2}{r_+^{2d_2}}\bigg)\\&+m^2\bigg(\frac{a_1\beta}{d_2}r_+^{d_2}+a_2\beta^2r_+^{d_3}+d_3a_3\beta^3r_+^{d_4}+d_3d_4a_4\beta^4r_+^{d_5}\bigg)\bigg].\label{22}\end{split}
\end{eqnarray}
Note that $\Sigma_{d_2}$ labels the volume of $d_2$-dimensional hyper-surface. Calculation of Hawking temperature is crucial in order to gain insight into the thermodynamic behaviour of the dyonic black holes provided by Eq. (\ref{17}). Hence, one might utilize the terminology
\begin{equation}
T=\frac{\kappa_s}{2\pi}, \label{25}
\end{equation}
where the quantity $\kappa_s$ refers to the surface gravity and is specified as
\begin{eqnarray}
\label{26}
\begin{split} 
\kappa_s&=\sqrt{-\frac{1}{2}(\nabla_{\alpha}\mathcal{X}_{\beta})(\nabla^{\alpha}\mathcal{X}^{\beta})}. 
\end{split} 
\end{eqnarray}
Here $\mathcal{X}_{\alpha}$ symbolizes the time-like Killing vector field. Thereby, one may get to   
\begin{eqnarray}
\label{27}
\begin{split} 
T(r_+)&=\frac{1}{4\pi}\bigg[\frac{\Theta d_3}{r_+}-\frac{2\Lambda r_+}{d_2}-\frac{\nu^2}{2d_2r_+^{2d-5}}-\frac{e^2r_+}{2d_2\big(r_+^{2d_2}+8\chi\nu^2(\Gamma(d_1))^2\big)}\\&+m^2\bigg(a_1\beta+\frac{a_2\beta^2d_3}{r_+}+\frac{d_4d_3a_3\beta^3}{r_+^2}+\frac{d_3d_4d_5a_4\beta^4}{r_+^3}\bigg)\bigg]. 
\end{split} 
\end{eqnarray}
\begin{figure}[h]
	\centering
	\includegraphics[width=0.8\textwidth]{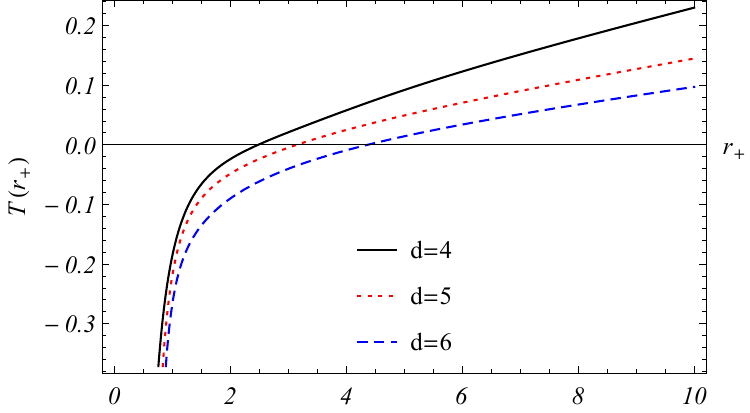}
	\caption{Behaviour of temperature (Eq. (\ref{27})) in various spacetime dimensions. The other parameters are fixed as $\Lambda=-0.3$, $\nu=2.5$, $e=10$, $m=0.1$, $\Theta=-1$, $a_1=1$, $a_2=2$, $a_3=1$, $a_4=2$, $\beta=1$, and $\chi=1$.}\label{skr2A}
\end{figure}
\begin{figure}[h]
	\centering
	\includegraphics[width=0.8\textwidth]{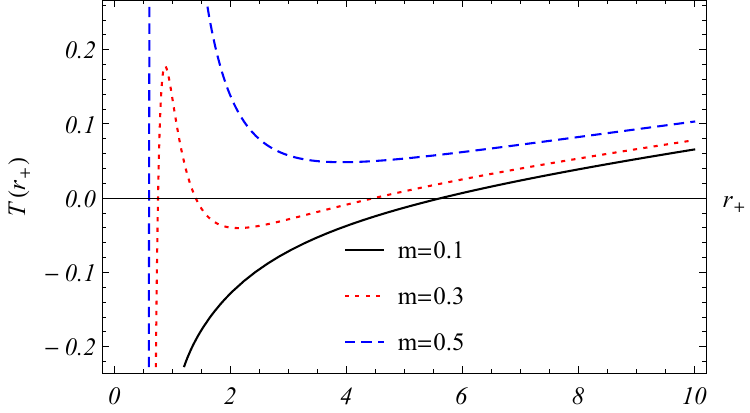}
	\caption{Dependence of $T(r_+)$ (Eq. (\ref{27})) on graviton's mass. The other parameters are fixed as $\Lambda=-0.3$, $\nu=2.5$, $e=10$, $d=7$, $\Theta=-1$, $a_1=1$, $a_2=2$, $a_3=1$, $a_4=2$, $\beta=1$, and $\chi=1$.}\label{skr2B}
\end{figure}
\begin{figure}[h]
	\centering
	\includegraphics[width=0.8\textwidth]{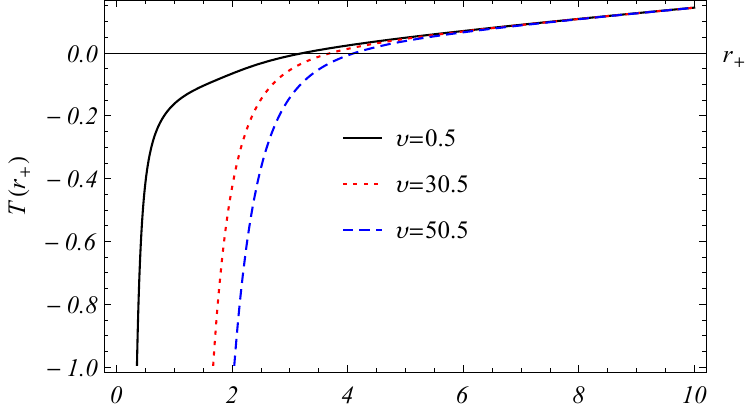}
	\caption{Plot of $T(r_+)$ (Eq. (\ref{27})) for various values of magnetic charge. The fixed values are chosen as $\Lambda=-0.3$, $m=0.05$, $e=10$, $d=5$, $\Theta=-1$, $a_1=1$, $a_2=2$, $a_3=1$, $a_4=2$, $\beta=1$, and $\chi=1$.}\label{skr2C}
\end{figure}
\begin{figure}[h]
	\centering
	\includegraphics[width=0.8\textwidth]{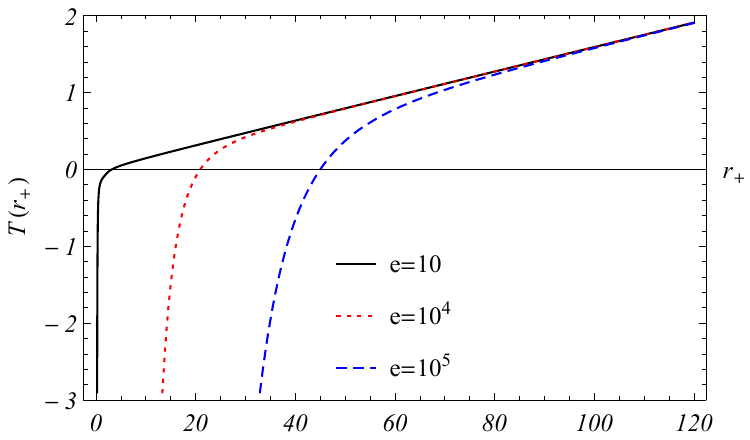}
	\caption{Impact of electric charge on $T(r_+)$ (Eq. (\ref{27})) for $\Lambda=-0.3$, $m=0.05$, $q=0.5$, $d=5$, $\Theta=-1$, $a_1=1$, $a_2=2$, $a_3=1$, $a_4=2$, $\beta=1$, and $\chi=1$.}\label{skr2D}
\end{figure} 
\textcolor{black}{Fig. \ref{skr2A} demonstrates that the event horizon $r_{ext}$ of extreme black hole (i.e. for which $T(r_{ext})=0$) is growing larger when the dimensions of spacetime rise.} Similarly, Fig. \ref{skr2B} illustrates how graviton's mass influences $T(r_+)$. Those values of $r_+$ at which $T(r_+)$ flips from negative to a positive sign correspond to first-order phase transition points. In addition, the positivity of $T(r_+)$ suggests the physicality of resultant dyonic black hole (\ref{17}). It is also observed that the four-dimensional dyonic black holes are more physical than the higher dimensional objects. Likewise, as the mass of graviton rises, the region concerning to physicality of objects widens. Moreover, one can witness the implications of the charges $\nu$ and $e$ on $T(r_+)$ in Figs. \ref{skr2C} and \ref{skr2D}, respectively. It is portrayed that the extreme black hole's horizon radius ascends and first-order phase transition points shift to right when the magnitudes of these charges go up. It is vital to remember that when $m=0$, the temperature of dyonic black hole in ETG is regained. Correspondingly, when $\nu$ and $e$ are vanishing, the Hawking temperature of the neutral black holes of massive gravity in diverse dimensions would be encountered. 

By utilizing the area law \cite{34,35,36}, the entropy of the dyonic black hole (\ref{17}) can be figured out as 
 \begin{eqnarray}\begin{split}
 S=\frac{\Sigma_{d_2}r_+^{d_2}}{4}.\label{28}\end{split}
 \end{eqnarray}
  The electric and magnetic charges can be estimated by employing the fluxes of $F_{[2]}$ and $H_{[d_2]}$ at infinity, respectively. Hence, one may introduce
  \begin{align}\begin{split}
  &e\sim \int\ast F_{[2]} ,\\&
  \nu\sim\int H_{[d_2]},\label{29}\end{split}
  \end{align} 
  with appropriate constants of proportionality. If $e$ and $\nu$ are considered as extensive thermodynamic variables then it is easier to showcase the first law as
  \begin{equation}
  dM=TdS+\Phi_e de+\Phi_{\nu}d\nu,\label{30}
  \end{equation} 
  where the conjugate quantities that correspond to $e$ and $\nu$ can be respectively presented as
  \begin{equation}
  \Phi_e=\frac{e\Sigma_{d_2}}{2d_2d_3r_+^{d_3}}F_1\bigg(\bigg[1,\frac{d_3}{2d_2}\bigg],\bigg[\frac{3d-7}{2d_2}\bigg],\frac{-8\chi \nu^2(\Gamma(d_1))^2}{r_+^{2d_2}}\bigg),\label{31}
  \end{equation}
  and
  \begin{equation}\begin{split}
  \Phi_{\nu}&=\frac{\nu^2\Sigma_{d_2}}{2d_3r_+^{d_3}}+\frac{e^2\Sigma_{d_2}r_+^{d_1}}{2d_2\nu\big(r_+^{2d_2}+8\chi\nu^2(\Gamma(d_1))^2\big)}\\&-\frac{e^2\Sigma_{d_2}}{2d_2\nu r_+^{d_3}}F_1\bigg(\bigg[1,\frac{d_3}{2d_2}\bigg],\bigg[\frac{3d-7}{2d_2}\bigg],\frac{-8\chi \nu^2(\Gamma(d_1))^2}{r_+^{2d_2}}\bigg).\label{32}\end{split}
  \end{equation}
 The heat capacity can be defined as
\begin{eqnarray}\begin{split}
C_H=T(r_+)\frac{\partial S}{\partial r_+}\bigg(\frac{\partial T}{\partial r_+}\bigg)^{-1}\bigg |_{e,\nu}.\label{33}\end{split}
\end{eqnarray}
Therefore, by plugging Eqs. (\ref{27}) and (\ref{28}) into Eq. (\ref{33}), one gets
\begin{eqnarray}\begin{split}
C_H&=\frac{d_2\Sigma_{d_2}r_+^{d_2}\big[\Theta d_2d_3r_+^2-2\Lambda r_+^4+\mathfrak{H}_1(r_+)\big]}{\big[\mathfrak{H}_2(r_+)-4\Theta d_2d_3r_+^2-8\Lambda r_+^4\big]},\label{34}\end{split}\end{eqnarray}
where
\begin{eqnarray}\begin{split}
\mathfrak{H}_1(r_+)&=d_2m^2\bigg(a_1\beta r_+^3+a_2\beta^2d_3r_+^2+d_3d_4a_3\beta^3r_+\\&+d_3d_4d_5a_4\beta^4\bigg)-\frac{\nu^2}{2r_+^{2d_4}}-\frac{e^2r_+^4}{2\big(r_+^{2d_2}+8\chi\nu^2(\Gamma(d_1))^2\big)},\label{35}\end{split}\end{eqnarray}
and
\begin{eqnarray}\begin{split}
\mathfrak{H}_2(r_+)&=\frac{(4d-10)\nu^2}{r_+^{2d_4}}+\frac{2e^2r_+^4\big((2d-3)r_+^{2d_2}-8\chi \nu^2(\Gamma(d_1))^2\big)}{\big(r_+^{2d_2}+8\chi\nu^2(\Gamma(d_1))^2\big)^2}\\&-4d_2m^2\bigg(a_2\beta^2d_3r_+^2+2d_3d_4a_3\beta^3r_++3d_3d_4d_5a_4\beta^4\bigg).\label{36}\end{split}\end{eqnarray}
\begin{figure}[h]
	\centering
	\includegraphics[width=0.8\textwidth]{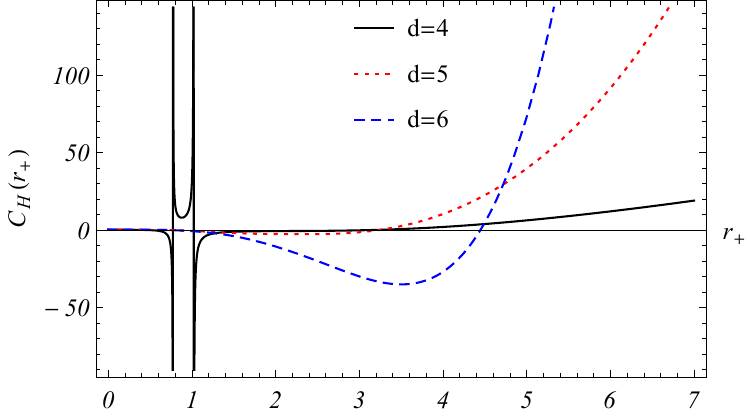}
	\caption{Behaviour of heat capacity (Eq. (\ref{34})) in various spacetime dimensions. The other parameters are fixed as $\Lambda=-0.3$, $\nu=0.5$, $e=10$, $m=0.05$, $\Theta=-1$, $a_1=1$, $a_2=2$, $a_3=1$, $a_4=2$, $\beta=1$, and $\chi=1$.}\label{skr4A}
\end{figure} 
\begin{figure}[h]
	\centering
	\includegraphics[width=0.8\textwidth]{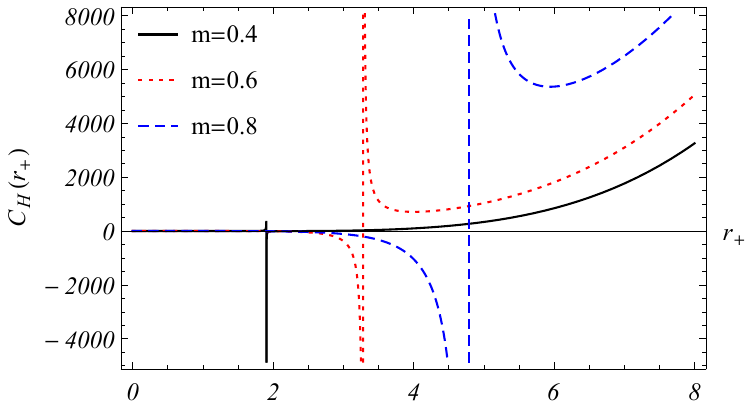}
	\caption{Dependence of heat capacity (Eq. (\ref{34})) on the graviton's mass various. The other parameters are fixed as $\Lambda=-0.3$, $\nu=0.5$, $e=10$, $d=6$, $\Theta=-1$, $a_1=1$, $a_2=2$, $a_3=1$, $a_4=2$, $\beta=1$, and $\chi=1$.}\label{skr4B}
\end{figure}
\begin{figure}[h]
	\centering
	\includegraphics[width=0.8\textwidth]{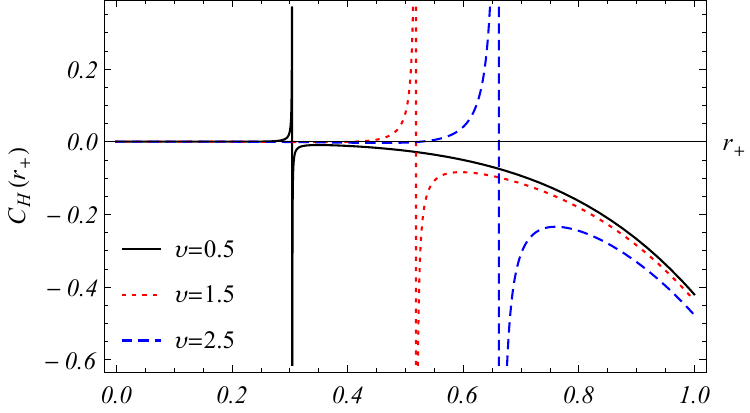}
	\caption{Plot of $C_H(r_+)$ (Eq. (\ref{34})) for various values of magnetic charge. Additionally, we have selected $\Lambda=-0.3$, $d=6$, $e=10$, $m=0.8$, $\Theta=-1$, $a_1=1$, $a_2=2$, $a_3=1$, $a_4=2$, $\beta=1$, and $\chi=1$.}\label{skr4C}
\end{figure}
\begin{figure}[h]
	\centering
	\includegraphics[width=0.8\textwidth]{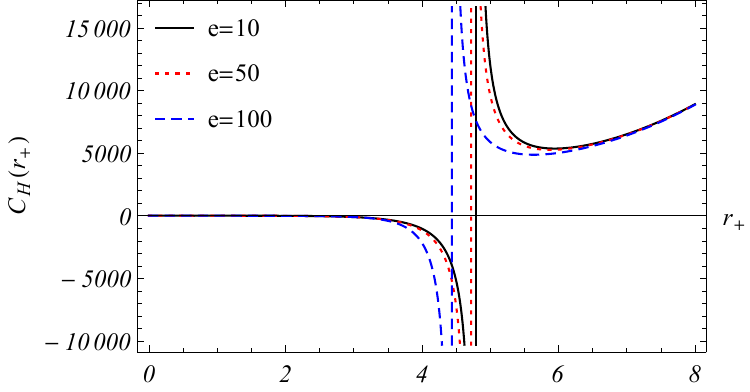}
	\caption{Plot of $C_H(r_+)$ (Eq. (\ref{34})) for various values of electric charge. The other parameters are fixed as $\Lambda=-0.3$, $d=6$, $\nu=2.5$, $m=0.8$, $\Theta=-1$, $a_1=1$, $a_2=2$, $a_3=1$, $a_4=2$, $\beta=1$, and $\chi=1$.}\label{skr4D}
\end{figure}

Figs. \ref{skr4A}, \ref{skr4B}, \ref{skr4C}, and \ref{skr4D} illustrate the plots of heat capacity for several values of the dimensionality parameter, graviton's mass, magnetic charge, and electric charge, respectively. The region of positive heat capacity conveys the local stability of the resultant dyonic black hole of massive gravity. The points at which the curves linked to $C_H(r_+)$ meet with the $r_+$-axis reflect the appearance of the first-order phase transitions, whereas the points at which this quantity is not regular reveal the emergence of second-order phase transitions. It is detected that heat capacity has two singular points, let us call them $r_1$ and $r_2$ such that $r_2>r_1$. The smaller black holes whose horizon radii falls in $(0,r_1)$ are stable. The black hole whose event horizon radius belong to the interval $(r_1,r_2)$, however, is locally unstable. Besides, the objects of larger sizes with horizon radii in the range $(r_2,\infty)$ are locally stable. Additionally, it has been recognized that when the magnitudes of electric and magnetic charges increase, $r_1$ grows while $r_2$ shrinks. It is also noteworthy to point out that the situation $m=0$ in Eq. (\ref{34}) refers to heat capacity of dyonic black holes in ETG, whereas by plugging $e=\nu=0$ in this equation leads to the heat capacity of neutral black holes in massive gravity.
\section{Thermal fluctuations}
Here, we are looking into how thermal fluctuations may affect the local thermodynamic stability of resultant dyonic black holes (\ref{17}). The effects of thermal fluctuations leads to the extended forms of various thermodynamic quantities, however, some quantities remain unaltered. The black hole entropy when encounters first-order correction becomes proportional to $\ln{C_VT^2}$ with $\bigg(C_V=\frac{1}{T^2}\frac{\partial^2S_1}{\partial\zeta^2}\bigg)_{\zeta=\zeta_0}$ and $\zeta=1/T$ \cite{109}. Following \cite{110,111,112,113,114}, one may express
 \begin{equation}
 S_0=\frac{1}{T^2}\bigg(\frac{\partial^2S_1}{\partial\zeta^2}\bigg)_{\zeta=\zeta_0}.\label{37}
 \end{equation}
 Remember that $S_0$ reflects the uncorrected entropy while $S_1$ signifies the modified entropy as a result of thermal fluctuations. This allows us to express the corrected entropy as
\begin{eqnarray}\begin{split}
S_1&=S_0-\frac{1}{2}\ln{\bigg(S_0T^2\bigg)}.\label{38}\end{split}
\end{eqnarray}
It is also beneficial to introduce a correction parameter $\xi$ in the second term of Eq. (\ref{38}). Thereby, the modified entropy can be stated as
\begin{eqnarray}\begin{split}
S_1&=S_0-\frac{\xi}{2}\ln{\bigg(S_0T^2\bigg)}.\label{39}\end{split}
\end{eqnarray}
Thus by using Eq. (\ref{28}) one can get
\begin{eqnarray}\begin{split}
S_1&=\frac{r_+^{d_2}\Sigma_{d_2}}{4}-\frac{\xi}{2}\log{\bigg(\frac{r_+^{d_2}\Sigma_{d_2}}{4}\bigg)}-\xi T(r_+),\label{40}\end{split}
\end{eqnarray}
in which $T(r_+)$ is the Hawking temperature of the dyonic black hole calculated in Eq. (\ref{27}). It is crucial to notice that one of those entities that logarithmic correction cannot change is temperature. The Helmholtz free energy is defined by the equation
\begin{eqnarray}\begin{split}
F&=-\int S_1dT=-\int\bigg[\frac{r_+^{d_2}\Sigma_{d_2}}{4}-\frac{\xi}{2}\log{\bigg(\frac{r_+^{d_2}\Sigma_{d_2}}{4}\bigg)}\bigg]\mathfrak{H}_3(r_+)dr_+,\label{41}\end{split}
\end{eqnarray}
where
\begin{eqnarray}\begin{split}
\mathfrak{H}_3(r_+)&=-\frac{\Theta d_3}{r_+^2}-\frac{2\Lambda}{d_2}-\frac{m^2}{r_+^4}\big(a_2\beta^2d_3r_+^2+2d_3d_4a_3\beta^3r_++3d_3d_4d_5a_4\beta^4\big)\\&+\frac{(2d-5)\nu^2}{2d_2r_+^{2d_2}}+\frac{e^2\big((2d-3)r_+^{2d_2}-8\chi \nu^2(\Gamma(d_1))^2\big)}{\big(r_+^{2d_2}+8\chi\nu^2(\Gamma(d_1))^2\big)^2}.\label{42}\end{split}
\end{eqnarray}
Likewise, the corrected mass is obtained as
\begin{eqnarray}\begin{split}
M_1&=F+TS_1=F+\frac{1}{4\pi}\bigg[\frac{\Theta d_3}{r_+}-\frac{2\Lambda r_+}{d_2}-\frac{e^2r_+}{2d_2\big(r_+^{2d_2}+8\chi\nu^2(\Gamma(d_1))^2\big)}\\&-\frac{\nu^2}{2d_2r_+^{2d-5}}+m^2\bigg(a_1\beta+\frac{a_2\beta^2d_3}{r_+}+\frac{d_4d_3a_3\beta^3}{r_+^2}+\frac{d_3d_4d_5a_4\beta^4}{r_+^3}\bigg)\bigg]\\&\times \bigg[\frac{r_+^{d_2}\Sigma_{d_2}}{4}-\frac{\xi}{2}\log{\bigg(\frac{r_+^{d_2}\Sigma_{d_2}}{4}\bigg)}-\xi T(r_+)\bigg].\label{43}\end{split}
\end{eqnarray}
Thereby, the corrected heat capacity can be obtained as follows
\begin{eqnarray}\begin{split}
C_1(r_+)=\frac{dM_1}{dT}=T\frac{dS_1}{dT}=\frac{\big(d_2\Sigma_{d_2}r_+^{d_2}-2\xi\big)}{d_2\Sigma_{d_2}r_+^{d_2}}C_H(r_+)-\xi,\label{44}\end{split}
\end{eqnarray}
where $C_H$ is defined by Eq. (\ref{34}).
\begin{figure}[h]
	\centering
	\includegraphics[width=0.8\textwidth]{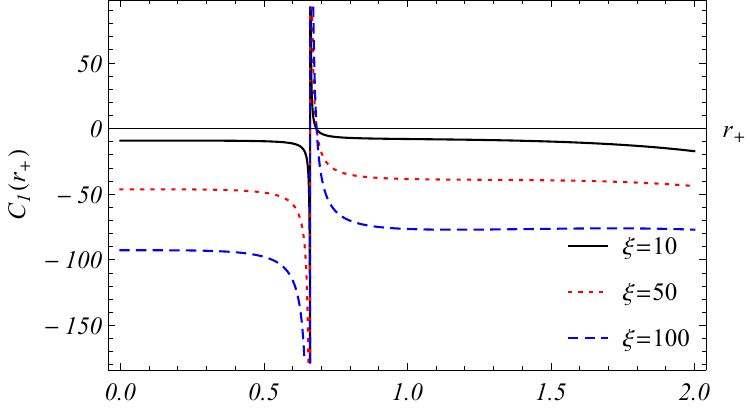}
	\caption{Plot of $C_H(r_+)$ (Eq. (\ref{34})) for various values of $\xi$. The rest of the parameters are chosen as follows: $\Lambda=-0.3$, $\Sigma_{d_2}=1$, $e=10$, $d=6$, $\nu=2.5$, $m=0.8$, $\Theta=-1$, $a_1=1$, $a_2=2$, $a_3=1$, $a_4=2$, $\beta=1$, and $\chi=1$.}\label{skr5}
\end{figure}
 Note that the substitution $\xi=0$ gives us the heat capacity $C_H$ from Eq. (\ref{44}). The impact of thermal fluctuation parameter on corrected heat capacity is pictured in Fig. \ref{skr5}. The stability of dyonic black holes provided by Eq. (\ref{17}) is reported to be seriously impacted by the logarithmic correction in entropy, while the phase transitions of the black holes are unaffected by thermal fluctuations.

 	\section{Shadow cast of dyonic black holes of massive gravity sourced by quasitopological electromagnetic field}\label{secs.Shadow}
Here, we explore the shadow cast by dyonic black holes that arise from massive gravity in four dimensions. These black holes are influenced by a quasitopological electromagnetic field.

The Lagrangian denoted as $\mathcal{L}(x,\dot{x})=\frac{g_{\mu\nu}}{2}\dot{x}^{\mu}\dot{x}^{\nu}$, which governs the paths taken by geodesics within a spacetime metric characterized by spherically symmetric and static attributes \cite{Cunha:2018acu}
	\begin{equation}
		\begin{aligned}\mathcal{L}(x,\dot{x})=\frac{1}{2}\left(f(r)\dot{t}^{2}-f(r)^{-1} \dot{r}^{2}-r^{2}\left(\dot{\theta}^{2}+\sin^{2}\theta\dot{\phi}^{2}\right)\right)~.\end{aligned}
	\end{equation}
	As is customary, employing the Euler-Lagrange equation $\frac{d}{d\lambda}\left(\frac{\partial\mathcal{L}}{\partial\dot{x}^{\mu}}\right)-\frac{\partial\mathcal{L}}{\partial x^{\mu}}=0$
	In the plane along the equator ($\theta=\pi/2$), two quantities remain conserved: the energy denoted as $E$ and the angular momentum represented by $L$ 
	\begin{equation} \label{cons}
		E=f(r)\dot{t},\quad L=r^{2}\dot{\phi}.
	\end{equation}
	Upon considering the equation governing null-geodesics for light, we arrive at:
	\begin{equation}\label{frG}
		f(r)\dot{t}^{2}-f(r)^{-1}\dot{r}^{2}+r^{2}\dot{\phi}^{2}=0\,.
	\end{equation} 
	Substituting the conserved quantities $E$ and $L$, as indicated in equation (\ref{cons}), into the aforementioned equation (\ref{frG}), yields the ensuing orbit equation for photons:
	\begin{equation}
		\left(\frac{dr}{d\phi}\right)^{2}=r^{2}f(r)\left(\frac{r^{2}}{f(r)}\frac{E^{2}}{L^{2}}-1\right)~.\label{eff}
	\end{equation}
	and the potential is written as
	\begin{equation}
		\left(\frac{dr}{d\phi}\right)^{2}=V_{eff}~,
	\end{equation}
	with 
	\begin{equation}
		V_{eff}=r^{4}\left(\frac{E^{2}}{L^{2}}-\frac{f(r)}{r^{2}}\right)~.
	\end{equation}
	Considering that the orbit equation exclusively relies on the impact parameter $b=L/E$ at the juncture where the trajectory shifts ($r=r_{ph}$), we are obliged to impose the criteria $dr/\left.d\phi\right|_{r_{ph}}=0$ or equivalently $V_{eff}=0,\quad V_{eff}^{\prime}=0$, as discussed in reference \cite{Perlick:2021aok,Khodadi:2022pqh}. This leads to the subsequent relationship for the impact parameter at the turning point:
	\begin{equation}
		b^{-2}=\frac{f(r_{ph})}{r_{ph}^{2}}\,.
		\label{impact}
	\end{equation}
	To determine the radius $r_{ph}$ of the photon sphere, it is necessary to enforce the requirements $dr/\left.d\phi\right|_{r_{ph}}=0$ and $d^{2}r/d\phi^{2}|_{r_{ph}}=0$. This results in the subsequent equations:
	\begin{equation} \label{photon0}
	\frac{d}{dr}(\frac{r^2}{f(r)})_{r_{ph}}=0~,
	\end{equation}
 
 	\begin{equation} \frac{f^{\prime}\left(r_{ph}\right)}{f\left(r_{ph}\right)}-\frac{2}{r_{ph}}=0.\label{photon}
\end{equation}
	By substituting equation (\ref{impact}) into equations (\ref{photon0}) and (\ref{photon}), one can straightforwardly ascertain the position of the photon sphere $r_{ph}$ and the pivotal impact parameter $b_{crit}$. Consequently, the structure of equation (\ref{eff}) can be reformulated as:
	\begin{equation}
		\left(\frac{dr}{d\phi}\right)^{2}=\left(\frac{r^4 f(r_{ph})}{r_{ph}^2 }-r^2f(r)\right).
	\end{equation}
	When computing the shadow radius $R_{sh}$ as perceived by an observer positioned at $r_0$, it is conventional to utilize the angle $\alpha_{\mathrm{sh}}$ formed between the trajectory of the light ray and the radial direction, in the subsequent manner \cite{Perlick:2021aok}:
	\begin{equation}
		\cot\theta_{\mathrm{sh}}=\left.\frac{1}{\sqrt{f(r)r^2}}\frac{dr}{d\phi}\right|_{r=r_{0}}.
	\end{equation}
	and

	\begin{equation}
		\cot^{2}\theta_{\mathrm{sh}}=\frac{r_0^2 f(r_{ph})}{r_{ph}^2f(r_0)}-1~,
	\end{equation}
	in which \footnote{$\sin^{2}\theta_{\mathrm{sh}}=(1+\cot^{2}\theta_{\mathrm{sh}})^{-1}$.} and using $b_{cr}$ of (\ref{impact}), gives
	\begin{equation}
		\sin^{2}\theta_{\mathrm{sh}}=\frac{b_{\mathrm{cr}}^{2}f\left(r_{0}\right)}{r_0^2}.
	\end{equation}


	The shadow radius of the black hole concerning a stationary observer at $r_{0}$ is given by:
	\begin{equation}
		R_{\mathrm{sh}}=r_{0}\sin\theta_{\mathrm{sh}}=\sqrt{\frac{r_{ph}^2f\left(r_{0}\right)}{f\left(r_{ph}\right)}}~,
	\end{equation}

 where for a static observer at far away distance reads as
 \begin{equation}
R_{\mathrm{sh}}=\frac{r_{p h}}{\sqrt{f\left(r_{p h}\right)}}
\end{equation}
Since in limit \(r_{0} \rightarrow \infty\), then \(f\left(r_{0}\right) \rightarrow 1\).

	\begin{figure}[htp]
	\centering
\includegraphics[width=0.8\textwidth]{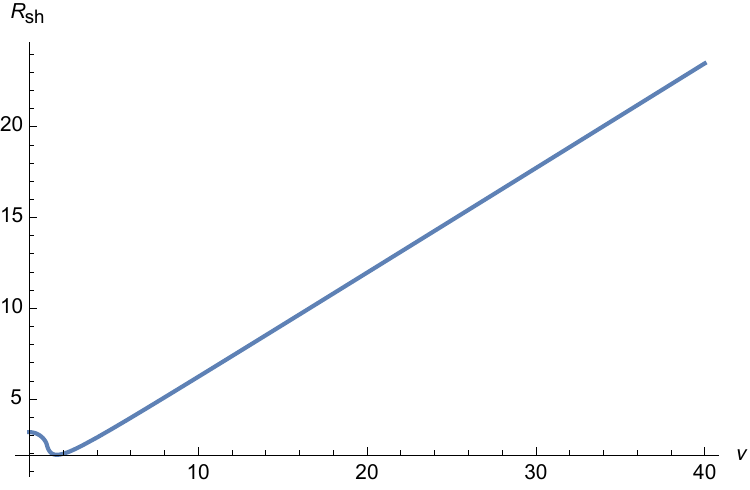}
	\caption{ Shadow cast by the black hole, with varying parameter $\nu$.  The rest of the parameters are chosen as follows: $\Lambda=0$,  $e=1$, $d=4$, , $m=0.8$,  $\mu=2$, $\Theta=-1$, $a_1=a_2=a_3=a_4=1$, $\beta=1$, and $\chi=1$.}\label{skr5a}
\end{figure}

	\begin{figure}[htp]
	\centering
\includegraphics[width=0.8\textwidth]{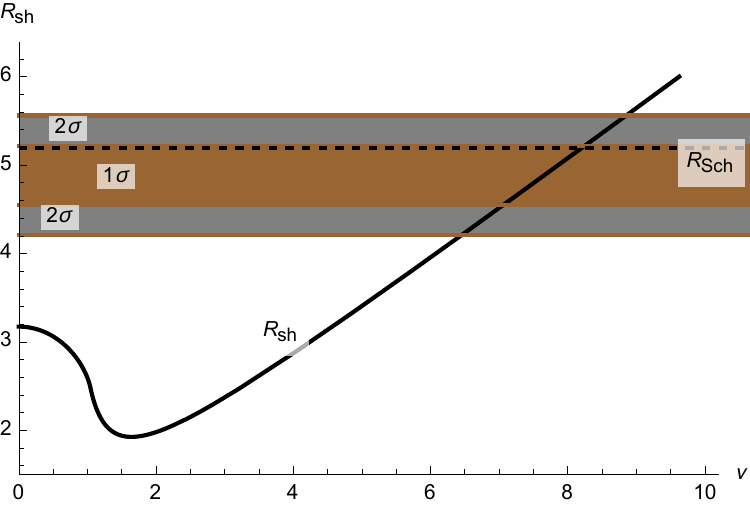}
	\caption{ The shaded regions in dark gray and light gray correspond to the areas consistent with the Event Horizon Telescope (EHT) horizon-scale image of Sgr A* at $1\sigma$ and $2\sigma$ confidence levels, respectively for $\Lambda=0$,  $e=1$, $d=4$, , $m=0.8$,  $\mu=2$, $\Theta=-1$, $a_1=a_2=0$, $a_3=a_4=1$, $\beta=1$, and $\chi=1$.}\label{fig16}
\end{figure}


We have now reached a point where we can effectively elucidate the influence of the underlying parameter $\nu$ on the dimensions of the shadow radius within a framework of spherical symmetry. In Figure \ref{skr5a}, we depict the shadows produced by varying values of $\nu$. The figure \ref{skr5a} shows the influence of the magnetic charge on the radius of the shadow.  It becomes clear that the parameter $\nu$ plays a pivotal role in diminishing the size of the shadow until $\nu$ reaches a value of 1.8. Beyond this threshold, a remarkable and rapid expansion of the shadow becomes evident.  
 It is clearly shown that the shadow radius decreases when $\nu$ is increased, up to a certain value, and then increases when $\nu$ is increased. Figure \ref{fig16} displays the variation of the shadow radius concerning the parameter $\nu$ in the black hole (solid curve) regimes. Additionally, the graph incorporates constraints derived from observational data, specifically the Event Horizon Telescope (EHT) image of Sgr A*.

\section{Summary and conclusion} 

In this study, we have investigated dyonic black holes of massive gravity sourced by quasitopological electromagnetic field in diverse dimensions. First, we have figured out the exact solution of the field equations and looked at how the graviton's mass, dimensionality parameter, and electromagnetic charges affected its geometrical properties. We have concluded that the solution (\ref{17}) describes dyonic black holes with inner and outer horizons for any non-negative value of the black hole's geometric mass such that $\mu\neq\mu_{ext}$. Correspondingly, when $\mu=\mu_{ext}$, the dyonic black hole has a single event horizon. Additionally, there also exists a critical value $m_c$ of the graviton's mass for which the solution (\ref{17}) has been demonstrated to describe black holes with inner and outer horizons for $m<m_c$, an extreme black hole for $m=m_c$, and naked singularities for $m>m_c$. We have also revealed that as the dimensionality parameter rises, the radii of inner and outer horizons grow. Meanwhile, when the magnitudes of charges $\nu$ and $e$ are on the rise, the inner horizon broadens while the outer horizon remains unaltered. Furthermore, the solution addressing $d$-dimensional dyonic black holes of ETG can be deduced if one plugs $m=0$ in Eq. (\ref{17}). Similarly, the neutral black holes of massive gravity would be recovered by putting $e=\nu=0$ in the resultant solution.

Next, we have examined thermodynamic properties of dyonic black holes and their local stability. It is concluded that based on the behaviour of temperature and selection of the graviton's mass, electric and magnetic charges, and the dimensionality parameter, the smaller black holes may not be physical while the larger black holes are physical. We also noticed that the horizon radii of extreme black holes are profoundly affected by variation of these parameters. Additionally, the first law of thermodynamics for the dyonic black holes of (\ref{17}) was also presented. Regarding the investigation of local thermodynamic stability, we realized that there exists two singular points $r_1$ and $r_2$ at which heat capacity is infinite. These divergences correlate to second-order phase transitions of black holes. We have encountered that the zones of local stability are $(0,r_1)$ and $(r_2,\infty)$. However, the black holes whose outer horizon radii belong to $(r_1,r_2)$ are locally unstable. In addition, the horizon of extreme black hole are shown to be greater in higher spacetime dimensions. It has also been noted that the charges $\nu$ and $e$ have a significant impact on the regions of local stability and phase transition points. For instance, as the magnitudes of electric and magnetic charges are on rise, $r_1$ grows while $r_2$ shrinks. We have additionally looked at how thermal fluctuations alter the thermodynamic quantities of the resultant dyonic black holes of massive gravity. By employing the logarithmic correction to the entropy of black hole, it became apparent that though the phase transition points do not depend on this correction, the zones of local stability are highly influenced by changing the values of $\xi$. Last, we are now poised to uncover the impact of the parameter $\nu$ on the spherical symmetry shadow size. In Figure 15, shadows are displayed for various $\nu$ values. Evidently, as $\nu$ increases, the shadow size initially decreases until reaching $\nu=1.8$ after which it rapidly expands.

Study of the phenomenas such as critical behaviour, quasi-normal modes, quantum evaporation and shadow cast of the resultant dyonic black holes could be attractive. One can also explore that how quasitopological electromagnetic field affects the physical properties of rotating black branes of massive gravity. Accordingly, physical characteristics of the black holes of other modified gravities within the framework of quasitopological electromagnetism may also be relevant. These concepts have been set aside for our upcoming work.

\acknowledgments 
A. {\"O}. would like to acknowledge the contribution of the COST Action CA21106 - COSMIC WISPers in the Dark Universe: Theory, astrophysics and experiments (CosmicWISPers) and the COST Action CA22113 - Fundamental challenges in theoretical physics (THEORY-CHALLENGES). We also thank TUBITAK and SCOAP3 for their support.

\end{document}